\documentclass[twocolumn]{aastex631}

\graphicspath{{./}{figures/}}
\usepackage[utf8]{inputenc}
\usepackage{graphicx}
\usepackage{dcolumn}
\usepackage{bm}
\usepackage{ulem} 
\usepackage{hyperref}

\usepackage{multirow,enumitem}
\usepackage{amsmath}
\usepackage{color}
\usepackage{xcolor}
\usepackage{multirow}

\newcommand{\LCDM}{\rm{\Lambda}CDM}

\begin{document}

\title{Determining Cosmological-model-independent $H_0$ with Gravitationally Lensed Supernova Refsdal}

\author{Xiaolei Li}
\affiliation{College of Physics, Hebei Normal University, Shijiazhuang 050024, China}
\author{Kai Liao}
\affiliation{School of Physics and Technology, Wuhan University, Wuhan 430072, China}

\correspondingauthor{Kai Liao}
\email{liaokai@whu.edu.cn}

\begin{abstract}

{{The reappearance of supernova Refsdal with detailed modeling of the lens cluster allows us to measure the time-delay distance,}} which serves as a powerful tool to determine the Hubble constant ($H_0$). We give a cosmological-model-independent method to estimate $H_0$ through Gaussian process regression, using time-delay measurements from this lensed supernova in combination with supernova data from the Pantheon+ sample. {{Using eight mass models for the lens cluster,}} we infer $H_0 = 64.2^{+4.4}_{-4.3} \, \rm{km\,s^{-1}\,Mpc^{-1}}$ and using two cluster models most consistent with the observations, we infer $H_0 = 66.3^{+3.8}_{-3.6} \, \rm{km\,s^{-1}\,Mpc^{-1}}$. Our estimates of the value of $H_0$ are in $1\sigma$ agreement with the results assuming a flat $\Lambda$CDM model and the uncertainties are comparable. Our constraint results on $H_0$ from the eight lens models and the two lens models indicate $2\sigma$ and $1.8\sigma$ tensions with that estimated by Supernova H0 for the Equation of State, respectively. However, our median values of $H_0$ from the two sets of lens models show good consistency with $H_0$ inferred from Planck cosmic microwave background observations assuming $\LCDM$ model within $1\sigma$. We also find that our results for $H_0$ indicate $2\sigma$ deviations and $1.7\sigma$ deviations from the constraint results of $H_0$ using six time-delay quasars by H0LiCOW with the same analysis method.  
\end{abstract}


\section{Introduction} \label{sec:intro}

The Hubble constant ($H_0$) describes the present expansion rate of the Universe, defined as $H_0 \equiv \Dot{a}/a$, when the scale factor $a = 1$ or $z = 0$, where $\Dot{a} = {\rm{d}}a/{\rm{d}}t$. Currently, there is a significant tension about the Hubble constant arising from the discrepancy between the local measurement of $H_0$ by the {the Supernova H0 for the Equation of State} (SH0ES) collaboration \citep{riess20162,riess2018type,riess2019large,Riess:2021jrx}, and the estimate of $H_0$ {{at early-Universe}} using {Planck} cosmic microwave background (CMB) and other cosmological observations assuming $\LCDM$ model \citep{ade2016planck,aghanim2018planck}. The tension has reached $> 5 \sigma$ level, indicating that either there exist unknown systematic errors in the observations or there are cosmological models beyond the standard $\LCDM$ model.

{{The time delays from strong gravitational lensing systems 
provide a part-directly method to determine $H_0$ (time delays are proportional to 1/$H_0$) and one can obtain $H_0$ assuming certain cosmological models \citep{Suyu:2009by,2019MNRAS.489.2097B,Millon:2019slk,Wong:2019kwg,Grillo:2020yvj,Grillo:2024rhi}.}}
{Time-delay distance $D_{\Delta t}$ could be obtained with detailed modeling of the potential well of the lens.} $D_{\Delta t}$ is a combination of three angular diameter distances and depends on $H_0$ and other cosmological parameters. Thus time-delay lensing could be used to determine $H_0$. Currently, time-delay cosmography of lensed quasars has achieved percent-level precision on the measurements of $H_0$ \citep{Birrer:2020tax,Millon:2019slk}. 

Strongly lensed supernovae (SLSNe) could provide time-delay measurements with comparative precision to lensed quasars but requiring shorter observation periods of $\sim$ months~\citep{Oguri:2002ku,Dobler:2006wv,Goobar:2016uuf,Oguri:2019fix,2022ChPhL..39k9801L,2023arXiv230107729S}. In 2014, the Hubble Space Telescope (HST) observed four images of a single supernova (SN) forming an Einstein cross configuration around an elliptical galaxy at redshift $z = 0.54$ in the MACS J1149.5+2223 ($11^{\text{h}}49^{\text{m}}35.8^{\text{s}},\,\, 22^{\circ}23'55''$ (J2000); hereafter MACS J1149) cluster. The cluster’s gravitational potential also creates multiple images of the spiral SN host galaxy at $z = 1.49$. Lens models of the galaxy cluster predicted that an additional image of the SN would be observed in 2015 \citep{Kelly:2014mwa}, which was detected by HST on 2015 December 11 UT \citep{Kelly:2015xvu}. The reappearance of SN was used to estimate the relative time delays and performed a blinded measurement of $H_0$ \citep{Kelly:2023mgv}. The authors inferred $H_0\,=\,64.8^{+4.4}_{-4.3}{\rm{km/s/Mpc}}$ using eight cluster lens models and $H_0\,=\,66.6^{+4.1}_{-3.3}\, {\rm{km/s/Mpc}}$ using two cluster lens models most consistent with observations.

However, in the work of \cite{Kelly:2023mgv}, the authors fixed the matter density $\Omega_{{M}}\,=\,0.3$ assuming a flat $\LCDM$ model to infer the value of $H_0$ with time-delay measurements from the lensed SN.  They claimed that the effect of the uncertainty from fixing $\Omega_{{M}}$ and $\Omega_\Lambda$ is smaller than $\sim 0.7 \,{\rm{km\,s^{-1}\,Mpc^{-1}}}$ given the weak dependence of $H_0$ on $\Omega_{{M}}$ and $\Omega_{\Lambda}$. Note that the key problem on $H_0$ is related to whether the flat $\LCDM$ model is the correct description. In a non-$\LCDM$ model framework, the bias could be unforeseeable. Another point to be concerned with is that, their estimates of $H_0$ are consistent with $H_0\,=\,   67.36 \pm 0.54\, {\rm{km\,s^{-1}\,Mpc^{-1}}}$ which was obtained from Planck CMB observations assuming $\LCDM$ model \citep{aghanim2018planck} while in tension with the latest local measurements from SH0ES which gave $H_0\, = \,73.04 \pm 1.04\, {\rm{km\,s^{-1}\,Mpc^{-1}}}$ \citep{Riess:2021jrx}. Therefore, it is worth double-checking the inferred $H_0$ value from different viewpoints. 

In this work, we are going to constrain the value of $H_0$ cosmological-model-independently based on Gaussian process (GP) regression. By using GP regression, we generate posterior samples of unanchored SN distance cosmological-model-independently and anchor them with the lensed SN system. 

This paper is organized as follows: in Section~\ref{sec:method}, we briefly introduce the data and the analysis method. Our results and discussions are presented in Section~\ref{sec:results}. Finally, we conclude in Section~\ref{sec:con}.


\section{Data and Methodology} \label{sec:method}
In our analysis, we use the time-delay measurements from the reappearance of SN Refsdal in combination with the SN data set from Pantheon+ \citep{Scolnic:2021amr}.

The reappearance of SN Refsdal, designed as SX was detected in HST observations taken on 2015 December 11 UT and the observation was used to estimate the relative time delay between SX and the previous image, S1, as 320-380 days. Given the matter distribution in the foreground MACS J1149 cluster lens, one can constrain the value of $H_0$ with time-delay cosmography.

In our work, we are going to use time-delay measurements of SN Refsdal considering two sets of lens models to make constraints on $H_0$ following the work in \cite{Kelly:2023mgv}, in which the authors first used a set of predictions from eight models, e.g., the Diego-a free-form model \citep{Treu:2015poa}, the Zitrin-c* light-trace-mass (LTM) model \citep{Treu:2015poa}, and the Grillo-g \citep{Grillo:2015dwl,Treu:2015poa}, Oguri-a* and Oguri-g* \citep{Treu:2015poa}, Jauzac-15.2* \citep{Jauzac:2015wfa} and Sharon-a* and Sharon-g* \citep{Treu:2015poa} parametric models. In the second, the authors only used the Grillo-g and Oguri-a* models. 

{Time-delay strong lensing gives the time-delay distance ($D_{\Delta t}$) measurement which is lens-model-dependent and cosmological-model-independent ($D_{\Delta t}$ measured in \cite{Kelly:2023mgv} did not rely on the assumption of $\Lambda$CDM model and the parameters therein. Since Kelly et al. 2023 fixed $\Omega_M$ in a flat $\Lambda$CDM model, which has only two parameters, the posterior of $H_0$ in their work can be transferred to the $D_{\Delta t}$ distribution used in our work. We do not give a new $D_{\Delta t}$ estimate in this work. We use the same $D_{\Delta t}$ posterior to anchor SNe Ia rather than assuming a $\LCDM$ model. Following~\cite{2019ApJ...886L..23L,Liao:2020zko}, we use unanchored SN data to get relative distances/unanchored distances with the GP algorithm, then use absolute distance $D_{\Delta t}$ to anchor them and get cosmological-model-independent $H_0$.}

 For the SN data, we use the so far largest data set, the Pantheon+ sample \citep{Scolnic:2021amr}, to provide an unanchored cosmological distance. {This sample consists of 1701 light curves of 1550 distinct SNe Ia ranging in redshift from $z = 0.001$ to 2.261. This larger SN Ia sample is a significant increase compared to the original Pantheon sample, especially at lower redshifts.}

With Pantheon+ SN data sets, we generate 1000 posterior samples of $D_{\rm{L}}H_0$ independently of the cosmological model using GP regression \citep{Holsclaw:2010nb,Holsclaw:2010sk,Shafieloo2012Gaussian,2023JCAP...02..014H}. The GP regression used here is based on the GPhist code \citep{2021ascl.soft09023K} first used in \cite{Joudaki:2017zhq}. GP regression works by generating a random set of cosmological functions whose statics are characterized by a covariance function, in which the kernel plays an important role. Moreover, the mean function also plays an important role in GP regression and the final reconstruction results are not quite independent of the mean function; however, it has a modest effect on the final reconstruction results because the values of hyperparameters help to trace the deviations from the mean function \citep{Shafieloo2012Gaussian,2013PhRvD..87b3520S}. In our work, we use a squared-exponential kernel for the covariance function with
\begin{equation}\label{eq:kernel}
   <\varphi(s_i)\varphi(s_j)>\,=\,\sigma_f^2 \exp\left({-\frac{|s_i-s_j|^2}{2\ell^2}}\right)
\end{equation}
where $s_i=\ln (1+z_i)/\ln (1+z_{\rm{max}})$ and $z_{\rm{max}}=2.261$ is the maximum redshift of the SN Ia sample.  $\sigma_f$ and $\ell$ are two hyperparameters that are marginalized over.
$\varphi$ is just a random function drawn from the distribution defined by the covariance function of Equation~(\ref{eq:kernel}) and we take this function as $\varphi(z)=\ln \left(H^{\rm{mf}}(z)/H(z)  \right)$, where $H^{\rm{mf}}(z)$ is the mean function, which we choose to be the best-fit $\LCDM$ model from Pantheon+ dataset. 

For the details of the reconstruction with GP, we refer the readers to \cite{Rasmussen:2006,Holsclaw:2010nb,Holsclaw:2010sk,Shafieloo2012Gaussian,2023JCAP...02..014H} and reference therein. 

Then, following ~\cite{2019ApJ...886L..23L}, we convert the 1000 posterior samples of $D_{\rm{L}}H_0$ to unanchored angular diameter distance $D_{\rm{A}}H_0$ with
\begin{equation}
    D_{\rm{A}}H_0=\frac{D_{\rm{L}}H_0}{(1+z)^2}.
\end{equation}

After that, we evaluate the values of 1000 $D_{\rm{A}}H_0$ at the redshift of the lens galaxy, $z_{\rm{d}}\,=\,0.54$, and the SN Refsdal's host galaxy, $z_{\rm{s}}\,=\,1.49$, to get 1000 values of $D_{\Delta t} H_0$ with
    \begin{equation}
       H_0D_{\Delta t}(z_d,z_s)\,=\,(1+z_{\rm{d}})\frac{(H_0D_{\rm{d}})(H_0D_{\rm{s}})}{(H_0D_{\rm{ds}})}.
    \end{equation}

In the end, we compute the likelihood of $H_0$ from each of the 1000 realizations in combination with the time-delay measurements, $D_{\Delta t}$ and then, marginalize over the realizations to form the posterior distribution of $H_0$. 

We need to note that, throughout this work, a spatially flat Universe is assumed. Thus, the angular diameter distance between the lens and the source is calculated via
\begin{equation}
    D_{\rm{ds}} \,=\,D_{\rm{s}}-\frac{1+z_{\rm{d}}}{1+z_{\rm{s}}}D_{\rm{d}}.
\end{equation}

\section{Results and Discussions} \label{sec:results}

\begin{figure*}
\centering
\includegraphics[width=0.45\textwidth]{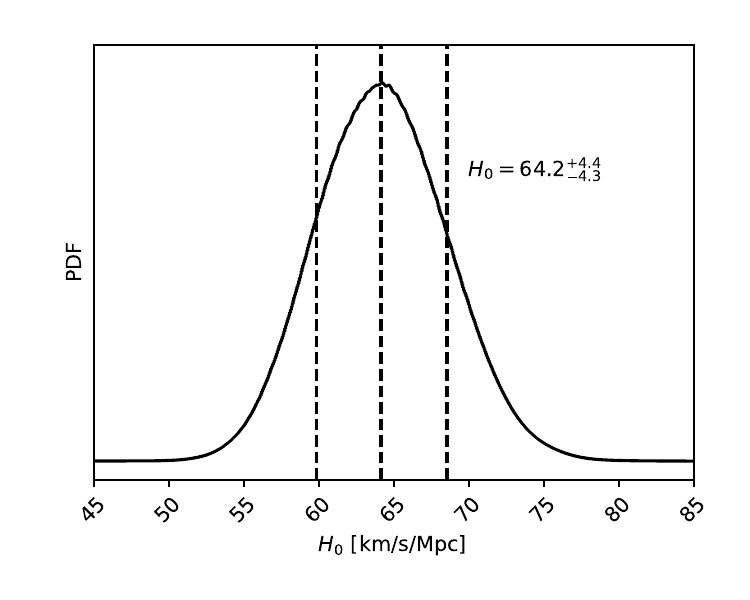}
\includegraphics[width=0.45\textwidth]{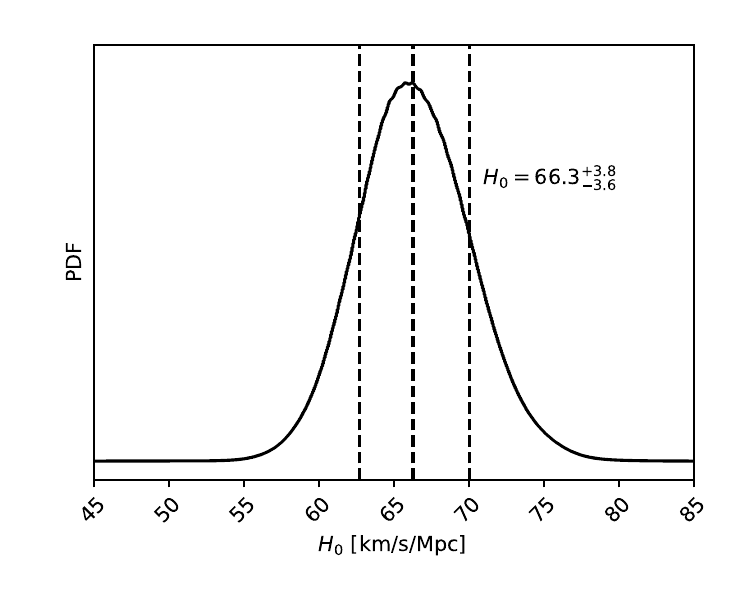}

\caption{The posteriors of $H_0$ from Pantheon+ sample in combination with SN Refsdal. The left plot is obtained with the full set of eight models constructed before the reappearance while the right is obtained with two models: Grillo-g and Oguri-a* models that are most consistent with observations.}
\label{fig:H0_posterior}
\end{figure*}
 With an unanchored cosmological distance from the SNe Ia anchored by a time-delay distance from SN Refsdal using GP regression, the posteriors of $H_0$ are shown in Figure~\ref{fig:H0_posterior}. For the cluster models of SN Refsdal, two sets of models are considered. The left plot is obtained with the full set of eight models constructed before the reappearance while the right plot is obtained with two models: Grillo-g and Oguri-a* models that are mostly consistent with the observations. Our cosmological-model-independent constraints give $H_0 = 64.2^{+4.4}_{-4.3} \, \rm{km\,s^{-1}\,Mpc^{-1}}$ ({median values and the 1$\sigma$ uncertainties}) for the eight cluster models for the lens cluster and $H_0 = 66.3^{+3.8}_{-3.6} \, \rm{km\,s^{-1}\,Mpc^{-1}}$ for the two models, respectively. 

Our two estimates of $H_0$ are within $1 \sigma$ agreement with
the results in \cite{Kelly:2023mgv}, in which the authors assumed a flat $\LCDM$ model with $\Omega_{\rm{M}}$ fixed. This consistency confirmed that the real cosmological evolution history is close to that $\Lambda$CDM describes. 
Our constraint results on $H_0$ from the eight lens models and the two lens models indicate $2\sigma$ and $1.8\sigma$ tension with that estimated by SH0ES in \cite{Riess:2021jrx}, respectively. However, our median values of $H_0$ from the two sets of lens models show good consistency with $H_0$ inferred from Planck CMB observation assuming a $\LCDM$ model within $1\sigma$ \citep{aghanim2018planck}. Furthermore, we realize that the uncertainties for our cosmological-model-independent analysis are comparable to the results assuming a specific cosmological model while reducing possible bias. The $H_0$ values from this lensed SN are also in tension with that obtained from six or seven time-delay measurements of lensed quasars, 
which give $H_0\,=\,73.3^{+1.7}_{-1.8}\,\rm{km\,s^{-1}\,Mpc^{-1}}$ in \cite{Wong:2019kwg}, $H_0\,=\,74.0^{+1.7}_{-1.8}\,\rm{km\,s^{-1}\,Mpc^{-1}}$ in \cite{Millon:2019slk} and $H_0\,=\,74.6^{+5.6}_{-6.1}\,\rm{km\,s^{-1}\,Mpc^{-1}}$ in \cite{Birrer:2020tax}. 

In \cite{Grillo:2024rhi}, 
the authors used the observed positions of 89 multiple images and 4 measured
time delays of SN Refsdal multiple images to measure the value of the Hubble constant, matter density, dark energy density, and dark energy density equation of state parameters. Without using any priors from other cosmological experiments, in an open wCDM cosmological model with cluster mass model based on \cite{Grillo:2015dwl,Grillo:2018ume,Grillo:2020yvj} (r model), 
the authors gave: $H_0=65.1^{+3.5}_{-3.4}$ km/s/Mpc, $\Omega_{\rm{M}}=0.76^{+0.15}_{-0.10}$ and $w = -0.92^{+0.15}_{-0.21}$ at the $68.3\%$ confidence level. The measured value of the Hubble constant is consistent with our results while having better precision.
More importantly, the author considered four cosmological models during the analysis and found the constraint results depend only very mildly on the underlying cosmological model and on the lens modeling details.

Comparing our constraint results to the time-delay cosmography using six time-delay quasars in combination with SN Ia with GP regression in \cite{Liao:2020zko}, we find our median results for $H_0$ are smaller than $H_0$ estimated from time-delay quasars by
$8.6\,\rm{km\,s^{-1}\,Mpc^{-1}}$ from the eight cluster models and $6.5\,\rm{km\,s^{-1}\,Mpc^{-1}}$ from the two cluster models, which correspond to $2\sigma$ deviations and $1.7\sigma$ deviations, respectively. 
{{The dominant source of uncertainty in the $H_0$ estimate is the lens model. If the lens galaxies or clusters can be perfectly modeled, it will provide a much more precise constraint on the value of $H_0$.}}

\begin{figure*}
\centering
\includegraphics[width=0.85\textwidth]{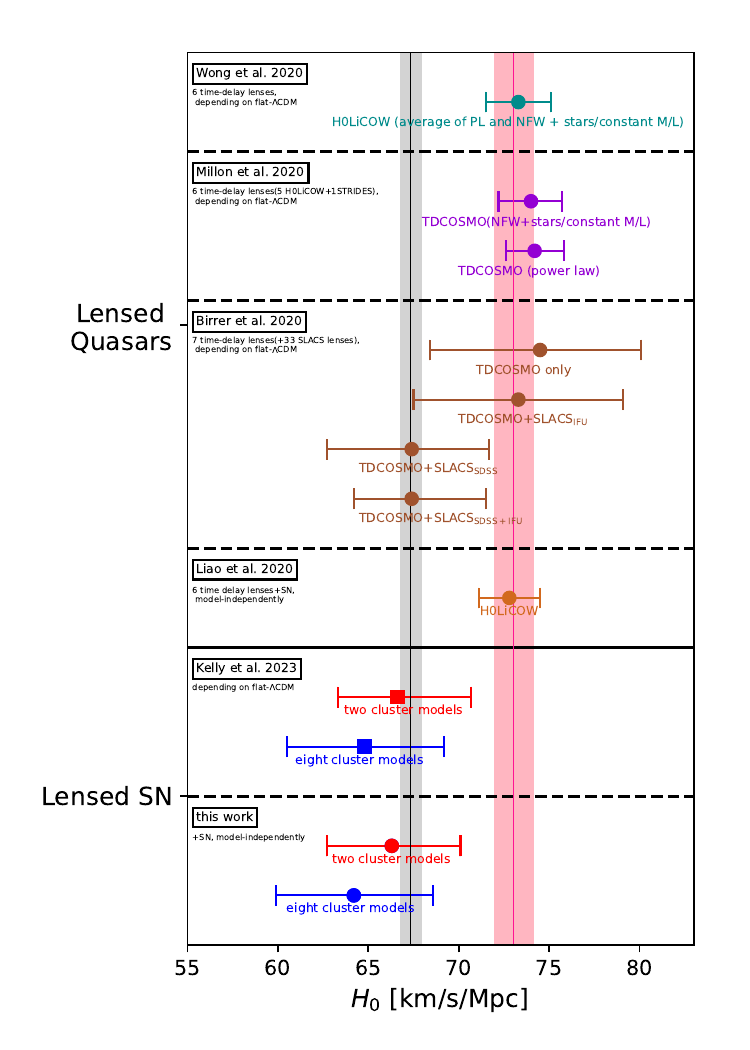}
\caption{Summary of the values of $H_0$ with $68\%$ CL constraints from time-delay lensing. Some of the results are adopted from \cite{Wong:2019kwg,Millon:2019slk,Birrer:2020tax,Liao:2020zko,Kelly:2023mgv}. The grey vertical band corresponds to the $H_0$ value from Planck CMB \citep{aghanim2018planck} and the pink vertical band corresponds to the $H_0$ values from SH0ES collaboration \citep{Riess:2021jrx}.}
\label{fig:H0values}
\end{figure*}

In the end, we show the values of $H_0$ from time-delay lensing measurements in Figure~\ref{fig:H0values} for direct comparison. The values of $H_0$ are adopted from \cite{Wong:2019kwg, Millon:2019slk, Birrer:2020tax,Liao:2020zko, Kelly:2023mgv, aghanim2018planck, Riess:2021jrx}.

\section{Conclusion} \label{sec:con}
We provide a cosmological-model-independent method using GP regression to estimate Hubble constant $H_0$ by anchoring the SNe Ia from the Pantheon+ sample with time-delay measurements of SN Refsdal assuming a spatially flat Universe. Using two sets of lens models, we obtained the constraints on $H_0$ of $64.2^{+4.4}_{-4.3} \, \rm{km\,s^{-1}\,Mpc^{-1}}$ and $ 66.3^{+3.8}_{-3.6} \, \rm{km\,s^{-1}\,Mpc^{-1}}$, respectively. 

Our results are within $1\sigma$ consistent with the results from \cite{Kelly:2023mgv} assuming a specific cosmological model. Despite not assuming a specific cosmological model, the uncertainties in our constraints are comparable to the results assuming $\LCDM$ model, while reducing possible bias toward the $\LCDM$ model. The results are in tension with that estimated by SH0ES in \cite{Riess:2021jrx} but agree well with $H_0$ inferred from Planck CMB observations assuming a $\LCDM$ model \citep{aghanim2018planck}.

We further compare our results to the $H_0$ estimate using six time-delay quasars using the same method described in our work \citep{Liao:2020zko}. The results show about $2\sigma$ deviations and $1.7\sigma$ deviations for the two sets of lens models, respectively. If we can model the lens cluster better in the future, we could obtain a much more precise estimate on the value of $H_0$.

For future estimates of the value of $H_0$, current surveys such as Vera C. Rubin Observatory's Legacy Survey of Space and Time, Euclid, and Nancy Grace Roman Space Telescope \citep{2010MNRAS.405.2579O,2015ApJ...811...20C,2015arXiv150303757S,2019arXiv190205569A}, will bring us over 100 lensed SNe Ia with well-measured time delays. Moreover, further analysis will help to improve cluster models and provide more precise constraints on $H_0$.

SN data will also be improved with ongoing and future surveys. Not only will the sample of the SNe become larger with better precision, but also the redshift coverage will become wider. With future time-delay distance measurements together with a larger and more precise SN Ia data set, one can obtain $H_0$ more precisely with the method described in this work and understand the $H_0$ tension better.


\section*{Acknowledgements}
\begin{acknowledgments}
We thank Patrick Kelly for providing the posteriors in their paper. This work was supported by National Natural Science Foundation of China (NSFC) Nos. 12222302, 12003006. X.Li is Funded by Science Research Project of Hebei Education Department No. BJK2024134.
\end{acknowledgments}

\bibliography{reference}{}
\bibliographystyle{aasjournal}

\end{document}